\documentclass[11pt]{article}
\usepackage{amsmath}
\usepackage{amssymb}
\usepackage{graphicx}

\def \aa{A\&A}

\def \aj{AJ}

\begin{document}

\setcounter{figure}{0}
\setcounter{table}{0}
\setcounter{footnote}{0}
\setcounter{equation}{0}

\vspace*{1.2cm}

\noindent {\Large EARTH ROTATION AND VARIATIONS OF THE GRAVITY FIELD, IN THE FRAMEWORK OF THE "DESCARTES-NUTATION" PROJECT}
\vspace*{1cm}

\noindent\hspace*{1.5cm} G. BOURDA                                         \\
\noindent\hspace*{1.5cm} SYRTE - UMR8630/CNRS, Observatoire de Paris       \\
\noindent\hspace*{1.5cm} 61 avenue de l'Observatoire - 75014 Paris, FRANCE \\
\noindent\hspace*{1.5cm} e-mail: Geraldine.Bourda@obspm.fr                 \\

\vspace*{2cm}



\noindent {\large 1. INTRODUCTION}

\smallskip

In this study, we investigate the links between Earth Orientation and Gravity Field Variations, on the basis of real gravity field data. The masses distributions inside the Earth govern the behaviour of the rotation axis in space (precession-nutation) and in the Earth (polar motion), as well as the Earth rotation rate (or equivalently, length of day). These distributions of masses can be measured by space owing to artificial satellites, the orbitography of which provides the Earth gravity field determination. Then, the temporal variations of the Earth gravity field can be related to the variations of the Earth Orientation Parameters (EOP) (with the Inertia Tensor). 

Nowadays, the Earth orientation measurements in space, obtained with Very Long Baseline Interferometry (VLBI), have a precision better than the milliarcsecond level. It is then necessary to consider all the geophysical sources that can improve the models precision. The goal of my PhD Thesis was to use the Earth gravity field measurements, as well as its variations, as a tool to improve the Earth orientation modelisation. We present here the theoretical point of view for such studies, as well as some results obtained.

\vspace*{1cm}


\noindent {\large 2. LINKS BETWEEN EARTH ORIENTATION PARAMETERS AND TEMPORAL VARIATIONS OF THE GRAVITY FIELD}

\smallskip

\begin{figure}[h]
\begin{minipage}[c]{.46\linewidth}
\centering
\includegraphics[width=5cm, angle=-90]{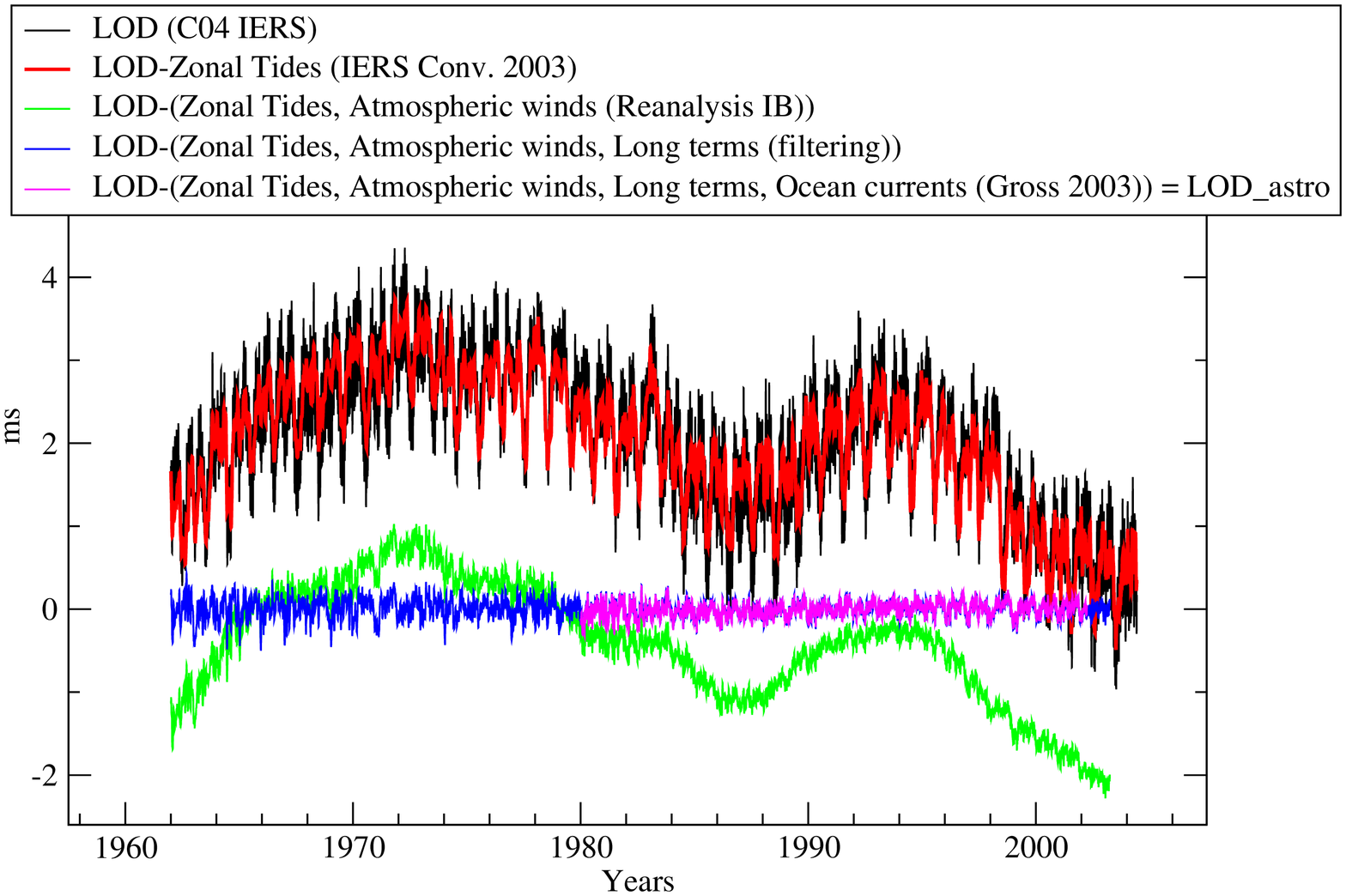}
\caption{Exces in the length of day: various components of the $\Delta(LOD)$ are removed.}
\label{fig:Bourda_fig1}
\end{minipage}
\hfill  
\begin{minipage}[c]{.46\linewidth}
\centering
\includegraphics[width=5cm, angle=-90]{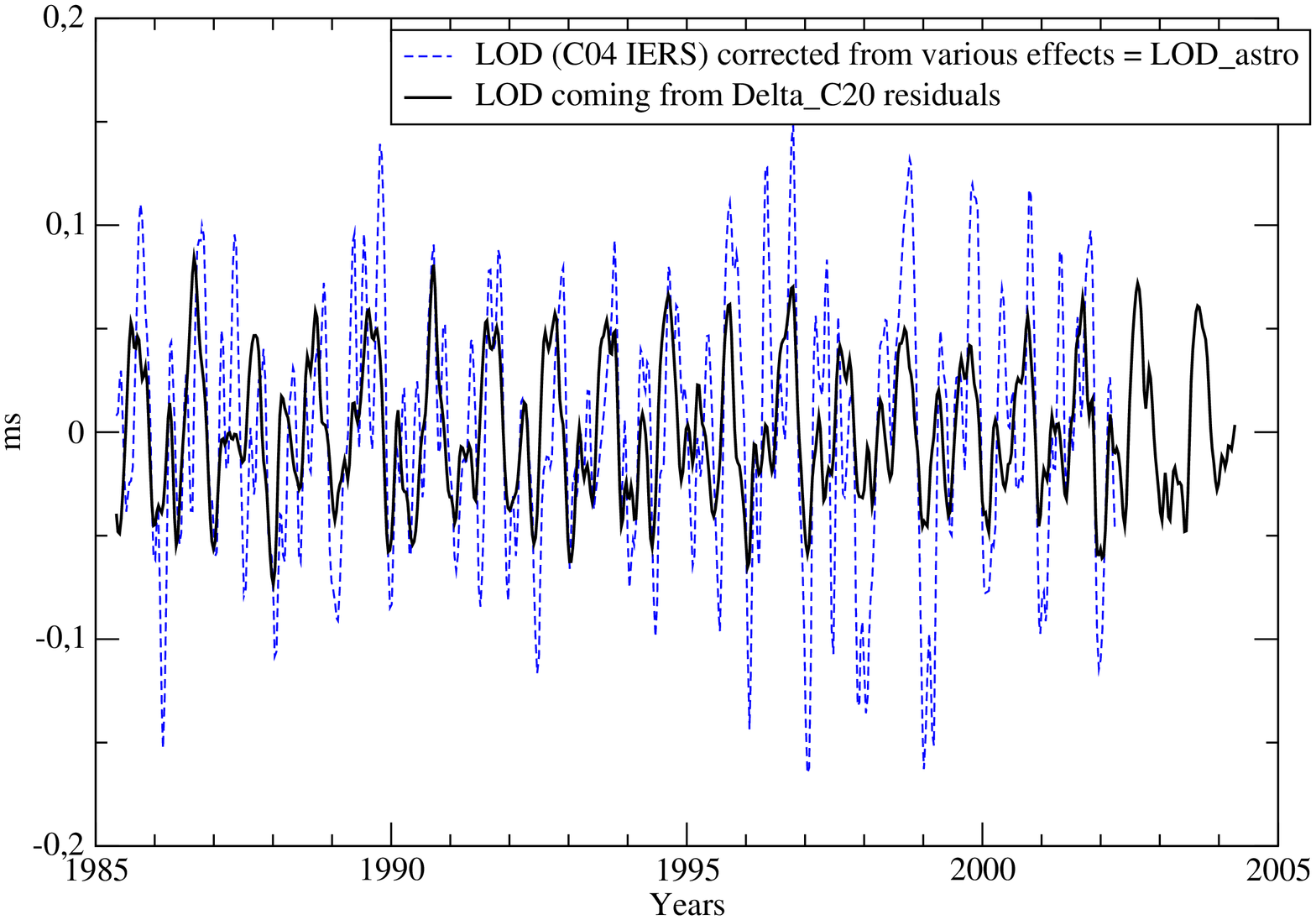}
\caption{$\Delta(LOD)$ obtained with observed $\Delta C_{20}$ data, and compared with $\Delta(LOD)_{astro}$ (see Fig.~\ref{fig:Bourda_fig1}).}
\label{fig:Bourda_fig2}
\end{minipage}
\end{figure}

The exces of length of day $\Delta(LOD)$ with respect to the nominal duration of 86400 s ($LOD_{\mbox{\footnotesize mean}}=86400$ s) can be related to the temporal variations of $C_{20}$ (i.e. $\Delta C_{20}$), coefficient of degree 2 and order 0 of the geopotential development into spherical harmonics. We obtain (see for example Gross~2000, Bourda~2003): 
\begin{equation}\label{eq:LOD}
\frac{\Delta(LOD)}{LOD_{\mbox{\footnotesize mean}}} = -\frac{2}{3 ~C_m} ~M ~{R_e}^2 ~\sqrt{5} ~\Delta \bar{C}_{20} + \frac{h_3}{C_m~\Omega}  
\end{equation}
where $C_m$ is the axial moment of inertia of the Earth mantle, $M$ and $R_e$ are respectively the mass and the equatorial radius of the Earth, $h_3$ is the axial relative angular momentum of the Earth, and $\Omega$ is the mean angular velocity of the Earth. The loading coefficient of 0.7 (Barnes~et~al.~1983), multiplying classically $\Delta \bar{C}_{20}$, is here a priori already considered into the $C_{20}$ real data. Figure \ref{fig:Bourda_fig2} shows the $\Delta(LOD)$ coming from the $\Delta C_{20}$ data of Biancale~\&~Lemoine~(2004), compared to the one of Fig.~\ref{fig:Bourda_fig1} coming from the C04 series of IERS where the "movement" effects and long terms have been removed. \\

The polar motion ($p=x_p - i~y_p$) excitation can be related to the geopotential coefficients of degree 2 and order 1 ($C_{21}$ and $S_{21}$), as following (see for example Gross~2000, Bourda~2003):
\begin{equation}\label{eq:polar_motion}
p + \frac{i}{\sigma_0}~\dot{p} = \dfrac{k_0}{k_0 - k_2} ~\dfrac{1}{C_m-A_m} \left[ \dfrac{h}{\Omega} - \sqrt{\dfrac{5}{3}} ~M ~{R_e}^2 ~(\bar{C}_{21}+i~\bar{S}_{21}) \right] 
\end{equation}
where $\sigma_0$ is the Chandler frequency, $k_0/(k_0 - k_2)=1.43$ (Barnes et al.~1983), $A_m$ is the equatorial moment of inertia of the mantle, and $h$ is the equatorial relative angular momentum of the Earth. We can notice that the loading coefficient of $(1+k'_2)$ (Barnes~et~al.~1983), multiplying classically $\bar{C}_{21}$ and $\bar{S}_{21}$, is here a priori already considered into the $C_{21}$ and $S_{21}$ real data. \\

An article about the link between the precession of the equator and the temporal variations of the $C_{20}$ geopotential coefficient have been publihed into \aa ~by Bourda~\&~Capitaine~(2004), on the basis of Williams~(1994) and Capitaine~et~al.~(2003). The $J_2$ rate influence onto the precession acceleration have been studied, as well as the $C_{20}$ periodic influence (annual and semi-annual terms), on the basis of the $\Delta C_{20}$ data of Biancale~et~al.~(2002).

\vspace*{1cm}


\noindent {\large 3. REFERENCES}
{

\leftskip=5mm
\parindent=-5mm
\smallskip

Barnes, R. T. H., Hide, R., White, A. A., and Wilson, C. A., 1983, {\it Proc. R. Soc. Lond.}, {\bf A 387}, pp. 31--73.

Biancale, R., Lemoine, J.-M., Loyer, S., Marty, J.-C., and Perosanz, F., 2002, private communication for the $C_{20}$ data coming from GRIM5.

Biancale, R., and Lemoine, J.-M., 2004, private communication for the $C_{20}$, $C_{21}$ and $S_{21}$ data coming from the redetermination of the gravity field with Lageos~I and II measurements.

Bourda, G., \ 2003, {\it Proceedings Journ\'{e}es 2002 Syst\`{e}mes de r\'{e}f\'{e}rence spatio-temporels}, edited by N.~Capitaine and M.~Stavinschi, Bucharest: Astronomical Institute of the Romanian Academy, Paris: Observatoire de Paris, p.~150-151.

Bourda, G., and Capitaine, N., 2004, \aa, {\bf 428}, pp. 691-702.

Capitaine, N., Wallace, P. T., and Chapront, J., 2003, \aa, {\bf 412}, pp. 567-586.

Gross, R. S., 2000, {\it Gravity, Geoid and Geodynamics 2000}, IAG Symposium 123, Sideris (eds.), Springer Verlag Berlin Heidelberg, pp. 153-158.

Williams, J. G., 1994, \aj, {\bf 108(2)}, pp. 711-724. 

}

\end{document}